\documentclass[12pt, a4paper]{article}

\usepackage{amsfonts}
\usepackage{amssymb}
\usepackage{amsmath}
\usepackage{array}
\usepackage{bm}
\usepackage{cancel}
\usepackage{cite}
\usepackage{epsfig}
\usepackage{graphicx, rotating}
\usepackage{hyperref}
\usepackage{latexsym}
\usepackage{multirow}
\usepackage{slashed}
\usepackage{verbatim}
\usepackage{xcolor}
\usepackage[export]{adjustbox}
\usepackage[normalem]{ulem}
\usepackage[utf8]{inputenc}

\newcommand{\be}{\begin{equation}}
\newcommand{\ee}{\end{equation}}
\newcommand{\bea}{\begin{eqnarray}}
\newcommand{\eea}{\end{eqnarray}}

\begin{document}

\begin{titlepage}

\begin{flushright}
\small
DESY-22-147
\\
IFT-UAM/CSIC-22-99
\end{flushright}
\vspace{.3in}

\begin{center}
{\Large\bf Tunneling Potential Actions}
\\[2mm]
{\Large\bf from Canonical Transformations}
\vskip 2mm
\bigskip\color{black}
\vspace{1cm}{
{\large
Jos\'e R.~Espinosa$^a$,
Ryusuke Jinno$^a$,
and
Thomas Konstandin$^b$
}}

{\small
\vskip 5mm
$^a$ Instituto de F\'{\i}sica Te\'orica IFT-UAM/CSIC\\ 
C/ Nicol\'as Cabrera 13-15, Campus de Cantoblanco, 28049, Madrid, Spain\\
\vskip 1mm
$^b$ Deutsches Elektronen-Synchrotron DESY, Notkestr.~85, 22607 Hamburg, Germany\\
}

\bigskip

\begin{abstract}
A new formulation for obtaining the tunneling action for vacuum decay based on the so-called tunneling potential was developed recently. In the original derivation, the new action 
was obtained by requiring that its variation led to the correct equations of motion and the same value for the action as evaluated on the Euclidean bounce. We present an alternative derivation of the tunneling potential action via canonical transformations. We discuss both single-field and multi-field cases as well as the case with gravity. We also comment on the possible application of the new approach to the calculation of the functional determinant prefactor in the tunneling rate.
\end{abstract}

\end{center}

\end{titlepage}

\newpage
\section{Introduction}
\label{sec:Introduction}

Tunneling is ubiquitous in condensed matter physics, cosmology, and elementary particle physics.
In quantum field theories this phenomenon occurs as the process of nucleation of a true vacuum bubble in the false vacuum, and the calculation of its rate has a long history.
The seminal paper~\cite{Coleman:1977py} presented a method to calculate the tunneling rate by evaluating the saddle-point configuration of the Euclidean action called the bounce.
The bounce is a configuration that stretches out into the true vacuum at the center while asymptotes to the false vacuum at far infinity.
It is a saddle-point configuration that has exactly one negative eigenvalue, which signals that the false vacuum is indeed unstable~\cite{Coleman:1987rm}.
The Euclidean approach was extended to the case with gravity in another seminal paper~\cite{Coleman:1980aw}.

Recently, a new formulation was proposed in~\cite{Espinosa:2018hue}.
This formulation calculates the tunneling action through a quantity called the tunneling potential $V_t(\phi)$.
One of the interesting aspects of this formulation is that the negative mode does not appear as it does in Euclidean calculations, so that the most probable decay path is realized as a minimum rather than a saddle point of the action.
This removes the computational difficulties associated with the negative mode, in particular when more than one scalar field are involved.
This new formulation was extended to multifield cases in~\cite{Espinosa:2018szu} and to the case with gravity in~\cite{Espinosa:2018voj}. 
The new approach has a number of other interesting features: 
It does not refer to the spacetime coordinate but completes within the field space alone;
it can also be used to generate potentials that admit analytic solutions and can easily be extended to the case of decay via thermal fluctuations, see~\cite{Espinosa:2018hue}.

The original derivation of the action for the tunneling potential 
was done by getting rid of Euclidean quantities, obtaining the equation of motion for $V_t$ and then deriving the action that leads to it.
The main goal of this paper is to explore the relation between the traditional Euclidean approach and the one with the tunneling potential and derive the latter action via a standard canonical transformation from the former.
Ultimately, relating these two formalisms in this manner might help to establish a way to compute the functional determinant prefactor in the tunneling potential formulation.

The organization of the paper is as follows.
In Sec.~\ref{sec:basic} we briefly review the tunneling-potential formulation.
In Sec.~\ref{sec:can} we show that the same action can be obtained in the $O(4)$ symmetric case through a canonical transformation for the single-field case without gravity.
In Sec.~\ref{sec:multi} we discuss multifield cases.
The derivation in the case with gravity is slightly trickier and is done in Sec.~\ref{sec:GR}.
In Sec.~\ref{sec:concl} we comment on the prospect for the calculation of the functional determinant prefactor and present some concluding remarks.

\section{Basic setup}
\label{sec:basic}

We are interested in the bounce solutions to the Euclidean action with $O(4)$ symmetry
\be
\label{eq:action1}
S_E = 2 \pi^2 \int_0^\infty 
\left[
\frac12 \left(\frac{d\phi}{dr}\right)^2 + V(\phi) \right] 
r^3 dr \, ,
\ee
that leads to the equation of motion
\be
\ddot \phi + \frac{3}{r} \dot \phi = V' \, .
\label{EEoM}
\ee
The dot denotes a derivative with respect to the 
space-time coordinate $r$, while the prime denotes a derivative with respect to
the field $\phi$.
The boundary conditions for the bounce are given by
\be
\dot\phi(0) = 0 \, , \quad
\phi(0) = \phi_0 \, , \quad
\phi(\infty) = \phi_+ \, ,
\ee
where $\phi_+$ denotes the metastable minimum (often the symmetric phase)
of the potential. The release point $\phi_0$ is 
varied to fulfil the boundary conditions and find the bounce.
In the following we assume $\phi_+ < \phi_0$.
The bounce is actually not a minimum of the 
action but a saddle point. This is no coincidence and signals 
the fact that the bounce represents a tunneling event~\cite{Coleman:1987rm}.

Alternatively, the action of the bounce can be found using the tunneling potential~\cite{Espinosa:2018hue}.
The tunneling potential is related to the bounce by
\be
\label{eq:def_Vt}
V_t = V(\phi) - \frac12 \left(\frac{d\phi}{dr}\right)^2 \ , 
\ee
where $d\phi/dr$ is evaluated for the bounce.
Notice that the equation of motion implies that the tunneling potential is monotonic
\be
\frac{dV_t}{dr} = \frac{3}{r} \left(\frac{d\phi}{dr}\right)^2 \geq 0 \, 
\ee
Likewise, the field $\phi$ and the space-time coordinate $r$ are monotonic and it follows that the tunneling potential can be considered as a (monotonic) function of the field value $\phi$.
With our convention $\phi_0>\phi_+$, one has $V_t'\leq 0$.

With the help of $V_t$ one can eliminate all Euclidean quantities to describe the bounce solution, using
\be
\dot\phi=-\sqrt{2(V-V_t)}\ ,\quad \ddot\phi =V'-V_t'\ , \quad
r=\frac{3\sqrt{2(V-V_t)}}{-V_t'}\ .
\label{r}
\ee
Taking a derivative of the last equation for $r$ with respect to $r$
an equation of motion for $V_t$ is obtained
\be
(4V_t'-3V')V_t'+6(V-V_t)V_t''=0\ ,
\ee
with boundary conditions
\be
V_t(\phi_+)=V(\phi_+)\ , \quad
V_t(\phi_0)=V(\phi_0)\ , \quad
V_t'(\phi_+)=0\ , \quad V_t'(\phi_0)=\frac34 V'(\phi_0)\ .
\ee
This equation of motion can be derived as the Euler-Lagrange equation from the following action
\be
\label{eq:action2}
S[V_t] = 54 \pi^2 \int_{\phi_+}^{\phi_0} \frac{(V-V_t)^2}{(-V_t')^3} d\phi \, ,
\ee
which also gives the correct value of the tunneling action on-shell.
Here, the tunneling potential $V_t$ and the potential $V$ are understood as functions 
of the scalar field $\phi$. The upper boundary of the integration is the
release point of the bounce that can be found by minimizing the action. 

This action has some interesting properties. First, the bounce is a minimum of this
action and not a saddle point. This simplifies the numerical search for the bounce action, 
in particular in the case with several scalar fields. Second, notice that the
action does not refer to the space-time coordinate $r$ in any way\footnote{Once 
$V_t(\phi)$ is obtained, the space-time coordinate $r$ can be reconstructed using (\ref{r}).}.
The action of the tunneling potential $V_t$ only operates in field space. This aspect will
be interesting again once gravity is included.

\section{Canonical transformation}
\label{sec:can}

In this section, we quickly review the generating function and the 
canonical transformation that can be used to transition from (\ref{eq:action1}) to (\ref{eq:action2}). 
Since the new action density (\ref{eq:action2}) is a function of the field $\phi$ but the coordinate $r$ is eliminated, to make the connection one has to interpret the action density (\ref{eq:action1}) in terms of $r(\phi)$ 
\be
S_E = \int_{\phi_+}^{\phi_0} L \, d\phi\ = -2 \pi^2 \int_{\phi_+}^{\phi_0} \left[ \frac1{2 r'} + r' \, V(\phi) \right] \, r^3 \, d\phi\ .
\ee
Next, we want to eliminate $r(\phi)$ in favor of the tunneling potential $V_t(\phi)$.

One way to achieve this is to use a generating function. To be precise, we will use a generating function $G(V_t,r)$
that depends on the old and new coordinates but not on the canonical momenta. We denote the canonical momentum of $r$ as $p$
and the canonical momentum of $V_t$ as $P$. The Hamiltonians are denoted as $H$ and $K$ correspondingly.

The Lagrangians of the old and new variables have then 
to coincide up to a total derivative, which is given by the generating function:
\be
p \, r' - H(r,p) = P \, V_t' - K(V_t,P) 
+ \frac{\partial G}{\partial \phi} + \frac{\partial G}{\partial r} r' + \frac{\partial G}{\partial V_t} V_t' \, .
\ee
This is achieved for 
\be
K = H + \frac{\partial G}{\partial \phi} \, , \quad 
p = \frac{\partial G}{\partial r} \, , \quad
P = -\frac{\partial G}{\partial V_t} \, .
\ee
We will use the second relation to reconstruct $G$ and then use the first and last relations to determine $K$ and $P$
and hence the Lagrangian in the new variables.

Notice that the $\partial G / \partial \phi$ term would give a boundary contribution $\left. G \right|$ to the action. In the cases
without gravity, $G$ will vanish at the start and end points of the bounce and this contribution is actually absent.
In the case with gravity, this contribution must be adjusted to reproduce the new action of the tunneling potential.

Furthermore, since we integrate the equation $p = \partial G / \partial r$ to find $G$, our choice 
will not be unique and we can always add a function of $V_t$ and $\phi$ to $G$ and change $P$ and the final action. Hence, such an additional term could in principle be used to simplify the new action of the tunneling potential. 

The canonical momentum is given by 
\be
p = \frac{\partial L}{\partial r'} =
2 \pi^2 \left[
\frac1{2 (r')^2} - V(\phi) \right] \, r^3 \, ,
\ee
and using the definition of $V_t$ this can be recast as
\be
p = - 2 \pi^2 V_t r^3 \, . 
\ee
Integration of the relation $p = \partial G / \partial r$ then yields
\be
G = - \frac12 \pi^2 V_t r^4 \, ,
\label{G0}
\ee
and\footnote{It can be checked that the Poisson bracket for the new variables $\left\{V_t,P\right\}=1$, as it should.}
\be
P = - \frac{\partial G}{\partial V_t} = \frac12 \pi^2 r^4 \, .
\ee
This relation and the definition of the tunneling potential (\ref{eq:def_Vt}) can then be inverted to 
express $r$ and $r'$ in terms of $V_t$ and $P$. For the new Hamiltonian, one finds 
\be
K = H + \frac{\partial G}{\partial \phi} = -4 \left(2 \pi^2 P^3\right)^{1/4} \sqrt{V(\phi) - V_t}\ \, .
\ee
Finally, the new Hamiltonian yields for the velocity of the 
tunneling potential
\be
V'_t = \frac{\partial K}{\partial P} = -3 \left(\frac{2 \pi^2}{P}\right)^{1/4} \sqrt{V(\phi) - V_t}\ ,
\ee
and the Lagrangian is found via a Legendre transformation, which reproduces (\ref{eq:action2}).

\section{Multiple scalar fields}
\label{sec:multi}

The tunneling potential approach with several scalar fields was previously studied in~\cite{Espinosa:2018szu}.
Also in this case, the action (the bold $\bm{\phi}$ is a vector in field space)
\be
\label{eq:action1m}
S_E = 2 \pi^2 \int_0^\infty 
\left[
\frac12 \left(\frac{d \bm{\phi}}{dr}\right)^2 + V(\bm{\phi}) \right] 
r^3 dr \, ,
\ee
can be reformulated as an action for the tunneling potential
\be
\label{eq:action2m}
S[V_t] = 54 \pi^2 \int_{\bm{\phi}_+}^{\bm{\phi}_0} \frac{(V-V_t)^2}{(-V_t')^3} d\varphi \, .
\ee
Here we introduced $\varphi$ to parametrize the length of the 
path, so that $\bm{\dot\phi} \cdot \bm{\dot\phi} = (\dot\varphi)^2$. Primes denote derivatives 
with respect to $\varphi$ while the dot is again the derivative with respect to $r$. 

The equation of motion is accordingly
\be
\bm{\ddot\phi} + \frac{3}{r}\bm{\dot\phi} = \bm{\nabla} V \, ,
\ee
formally very similar to the equation in the case with one field.

The proof that the action (\ref{eq:action2m})
reproduces the correct results is rather straight-forward but tedious. This is 
due to the fact that one has to split the scalar fields $\bm{\phi}$ into 
longitudinal and orthogonal parts (e.g.~using the Frenet-Serret basis) and 
the equations of motion look quite different for these two sets.

One possible trick to simplify the discussion is to use an additional 
path parameter ($\alpha$). This introduces an additional reparametrization invariance but makes the 
derivation again a bit more symmetric between $\varphi$ and the other degrees of freedom in $\bm{\phi}$.
The original action then reads 
\be
\label{eq:action1alpha}
S_E = 2 \pi^2 \int_0^\infty 
\left[
\left( \frac{d \bm{\phi}}{d \alpha} \cdot \frac{d \bm{\phi}}{d \alpha} \right)
\left(\frac{d r}{d \alpha}\right)^{-1}
+ V \, \frac{d r}{d \alpha} \right] \,
r^3 \, d\alpha\, ,
\ee
while the action for the tunneling potential turns into
\be
\label{eq:action2alpha}
S[V_t] = 54 \pi^2 \int_{\bm{\phi}_+}^{\bm{\phi}_0} \frac{(V-V_t)^2}{(-d V_t/d \alpha)^3} 
\left( \frac{d \bm{\phi}}{d \alpha} \cdot \frac{d \bm{\phi}}{d \alpha} \right)
d\alpha \, .
\ee
The equivalence of these two actions can then easily be shown by
replacing $r(\alpha)$ with $V_t(\alpha)$, defined by 
\be
V_t(\alpha) \equiv 
V - \frac12
\left( \frac{d \bm{\phi}}{d \alpha} \cdot \frac{d \bm{\phi}}{d \alpha} \right)
\left(\frac{d r}{d \alpha}\right)^{-2}
\ee
and following the same steps as before.

\section{Including gravity}
\label{sec:GR}

The case including gravity is also instructive to study~\cite{Espinosa:2018voj}. At first one might wonder how 
the action of the tunneling potential might look like. After all, there is no reference to
the space-time coordinate $r$ in the case without gravity while in gravity space-time is 
front and center. 

Using the $O(4)$ symmetric parametrization of the metric
\be
ds^2 = d\xi^2 + \rho(\xi)^2 d\Omega_3^2 \, ,
\ee
the conventional action (including a Gibbons-Hawking-York boundary term~\cite{York:1972sj, Gibbons:1976ue}) reads
\be
S_E = 2\pi^2 \int_0^{\xi_{\rm max}}
\left[ \rho^3 \left( \frac12 \dot \phi^2 + V\right)
- \frac{3\rho}{\kappa} (\dot\rho^2 + 1) \right] \, d\xi \, .
\label{SEgrav}
\ee
The dot denotes the derivative with respect to $\xi$, $\kappa$ is related to the reduced Planck mass, $\kappa = 1/m_P^2$, 
and $\xi_{\rm max}=\infty$ for Minkowksi or Anti de Sitter (AdS) false vacua while it is some finite constant for de Sitter (dS) vacua.
The two equations of motion then read
\be
\ddot \phi + \frac{3 \dot \rho}{\rho} \dot \phi - V' = 0 \, ,
\ee
and
\be
\ddot \rho = \frac{1}{2\rho}(1 - \dot \rho^2) - \frac12 \kappa \rho \left( V + \frac12 \dot \phi^2 \right) \, .
\ee
The full Einstein equations also imply the Friedmann equation
\be
\label{eq:Friedmann}
\dot \rho^2 = 1 + \frac{\kappa \rho^2}{3} \left( \frac12 \dot \phi^2 - V \right) \, ,
\ee
which automatically enforces the equation of motion for $\rho$.

\subsection{Non-positive cosmological constant}

When the cosmological constant is non-positive at the false vacuum, we have $\dot\rho>0$ and $\rho \to \infty$ for $\xi \to \infty$, 
which we assume in this subsection.

In order to use the canonical transformation, we write the action as a functional of 
$\xi(\phi)$ and $\rho(\phi)$. This yields 
\be
S_E = -2\pi^2 \int_{\phi_+}^{\phi_0}
\left[ \rho^3 \left( \frac1{2 \xi'^2} + V\right)
- \frac{3\rho}{\kappa} \left( \frac{\rho'^2}{\xi'^2} + 1 \right) \right] \, \xi' \, d\phi \, ,
\ee
where the prime again denotes derivatives with respect to $\phi$.

As a first step, let us consider the Hamiltonian of this system. The canonical momenta are found to be
\be
\label{eq:prho1}
p_\rho = \frac{\partial L}{\partial \rho'} = \frac{12 \pi^2 \rho \rho'}{\kappa \xi'} \, ,
\ee
(which is positive) and
\be
p_\xi = -\frac{\pi^2 \rho}{\kappa \xi'^2} 
\left[ 6 \rho'^2 - 6 \xi'^2 + \kappa \rho^2 (2 V \xi'^2 - 1) \right] \, .
\ee
Performing the Legendre transformation then yields the Hamiltonian 
\be
H = \frac{2 \pi^2 \rho^3}{\xi'} \, ,
\label{Hgrav}
\ee
which is surprisingly simple. 

Conventionally, the Hamiltonian would be expressed in terms of $\xi$, $\rho$ and 
the canonical momenta $p_\xi$ and $p_\rho$. This is in principle possible using these expressions, but 
notice that neither $p_\xi$ nor $p_\rho$ depend explicitly on $\xi$ but only on $\xi'$ such that also $H(\xi, \rho, p_\xi, p_\rho)$
will actually not depend on $\xi$.
This means that $p_\xi$ will be constant during the evolution.
Also notice that the momentum $p_\xi$ is proportional to the Friedmann equation (\ref{eq:Friedmann}), such that 
with the physical boundary conditions, we have $p_\xi=0$.
Using this information, any reference to $\xi$ vanishes in the Hamiltonian 
\be
H = -\frac{\rho}{\sqrt{6 \kappa}} \sqrt{p_\rho^2 \kappa^2 - 144 \pi^4 \rho^2 + 48 \pi^4 V \kappa \rho^4} \, ,
\ee
but notice that $H$ in principle depends explicitly on $\phi$ through the potential $V$.

Starting from this Hamiltonian, we would like to replace the variable $\rho$ by some appropriate $V_t$. We use the definition
\be
V_t = V- \frac{1}{2 \xi'^2} = -\frac{p_\rho^2 \kappa}{48 \pi^4 \rho^4} + \frac{3}{\kappa \rho^2} \, ,
\label{Vtgrav}
\ee
and express $p_\rho$ in terms of $V_t$ and $\rho$
\be
\label{eq:prho2}
p_\rho = \frac{12 \pi^2 \rho}{\kappa} \sqrt{1 - V_t \kappa \rho^2/3} \, .
\ee
Notice that for the bounce the square root is 
always real~\cite{Espinosa:2018voj} (see also below). 

This is readily integrated to obtain the generating function
\be
G = -\frac{12\pi^2 (1- V_t \kappa \rho^2/3)^{3/2}}{V_t \kappa^2} \, .
\ee
Notice that the generating function $G$ also does not depend on $V$ which will simplify the 
following steps. The canonical momentum gives\footnote{This can be rewritten, with $x\equiv 12\pi^2\rho/(\kappa p_\rho)$, as $P=(2\pi^2/3)\rho^4 x(1-3x^2)/(1-x^2)^2$ which, together with (\ref{Vtgrav}), leads immediately to $\left\{V_t,P\right\}=1$.}
\be
P = -\frac{\partial G}{\partial V_t} = -\frac{12\pi^2 (1 + V_t \kappa \rho^2/6)\sqrt{1 - V_t \kappa \rho^2/3}}{V_t^2 \kappa^2} \, ,
\ee
and the Hamiltonian reads
\be
K = -2 \pi^2 \rho^3 \sqrt{2(V - V_t)} \, .
\label{K}
\ee
Again, the Hamiltonian would have to be expressed in terms of $V_t$ and $P$, but
 this time this leads to a cubic equation and quite lengthy expressions.

Instead, we will proceed and perform all derivatives using the chain rule (keeping $V_t$ fixed), for example
\be
V_t' = \frac{\partial K}{\partial P} = \frac{3 K}{\rho} \left(\frac{\partial P}{\partial \rho}\right)^{-1}
= -\frac{3}{\rho} \sqrt{2(V-V_t)} \sqrt{1- V_t \kappa \rho^2/3} \, .
\ee
Notice that $V_t'$ is negative as it should be in our convention.

Using this relation, the new action of the tunneling potential can be 
constructed. The last remaining step is to express $\rho$ in terms of $V_t$ and $V_t'$.
Notice that this can be achieved via 
\bea
D^2 \equiv V_t'^2 + 6 \kappa (V - V_t) V_t= \frac{18(V-V_t)}{\rho^2} \, ,
\eea
such that the Lagrangian reads
\be
\label{eq:lagfinal1}
L = \frac{12 \pi^2 \left[ D^2 - 3 \kappa V_t (V - V_t) \right]}{D V_t^2 \kappa^2} \, .
\ee
This is not yet the final result, since two contributions are missing for the tunneling action. First, the total derivative
of $G$ contributes to the new action
\be
\label{eq:Gboundary}
\left. G \right |_+^0 = \left. -\frac{12\pi^2 (1- V_t \kappa \rho^2/3)^\frac32}{ V_t \kappa^2} \right|_+^0 \, ,
\ee
where here, and below, $0$ stands for $\phi=\phi_0$ or $\xi=0$ and $+ $ for $\phi=\phi_+$ or $\xi=\infty$.
Second, the false vacuum background contribution has to be subtracted from the action. In particular, this background contribution makes 
the action finite, that would otherwise diverge when integrating until infinity. This is analogous to normalizing the 
potential to zero in the symmetric phase in the case without gravity. In the case with gravity, this background
contribution has static $\phi=\phi_+$ but still non-trivial geometry. Using 
\be
\dot\rho = \sqrt{1 - V_+ \kappa \rho^2 /3} \, ,
\ee
one finds
\bea
S_{bg} &=& 4\pi^2 \int_0^+ \left( \rho^3 V_+ - \frac{3 \rho}{\kappa} \right) d\xi \, \\
&=& - \frac{12 \pi^2}{\kappa} \int_0^+ \, \rho \dot\rho \sqrt{1 - V_+ \kappa \rho^2 / 3} \ d\xi \\
&=& 
\left. \frac{12\pi^2 (1- V_+ \kappa \rho^2/3)^\frac32}{V_+ \kappa^2} \right|_0^+ \, .
\label{Sbg}
\eea
Notice that this expression coincides with $G$ in (\ref{eq:Gboundary}) but $V_t$ is replaced by $V_+$.
Both of these expressions are divergent when $\rho\to\infty$ but the divergence cancels in the difference
and one obtains
\be
\left. 
G \right|_+^0 - S_{bg} = \frac{12 \pi^2}{\kappa^2} \left( \frac{1}{V_+} - \frac{1}{V_0} \right) \, .
\ee
Adding the $\phi$-derivative of this contribution to the action density (\ref{eq:lagfinal1}) using 
\be
\int_+^0 d\phi \, \frac{V_t'}{V_t^2} = -\left. \frac{1}{V_t} \right|_+^0\, ,
\ee
one arrives at the final expression
\be
\label{eq:lagfinal2}
L = \frac{6\pi^2}{\kappa^2} \frac{(D + V_t')^2}{D V_t^2} \, ,
\ee
in agreement with the action found in~\cite{Espinosa:2018voj}.

\subsection{Positive cosmological constant}

The case of tunneling from de Sitter space with positive cosmological constant is somewhat special. The bounce action is finite due to the fact that for some finite $\xi_{\rm max}$ the space shrinks to zero, $\rho \to 0$, which marks the end point of the bounce. This implies, in particular, that at some point $\xi_\times$ the velocity $\dot \rho$ turns negative. Moreover, the bounce does not end in the symmetric phase $\phi_+$, since this is inconsistent with $\dot\phi=0$ and $\rho=0$. 

Nevertheless, when the cosmological constant is positive the derivation of the previous subsection equally applies but one has to pay special attention to the signs of several expressions. For the bounce, 
the sign of $\dot\rho$ will change at $\xi_\times$, and at the corresponding $\phi_\times\equiv\phi(\xi_\times)$ in turn $V_t'$ will change sign and $\sqrt{1 - V_t(\phi)\kappa \rho^2/3}$ will touch zero.
According to (\ref{eq:prho1}) also $p_\rho$ will change sign in this point and the correct expression instead of (\ref{eq:prho1}) is now
\be
\label{eq:prho3}
p_\rho = \frac{12 \pi^2 \rho}{\kappa} \sqrt{1 - V_t \kappa \rho^2/3} \,{\rm sign}{\dot\rho} \, .
\ee
The additional factor ${\rm sign}{\dot\rho}$ shows that there is a sign flip when $\dot{\rho}$ crosses zero but $p_\rho$ is continuous since the square root vanishes at that point.
Also $G$ will change sign at the same point, while $K$ and the Lagrangian (\ref{eq:lagfinal1}) are unchanged (they only contain $V_t'^2$ which does not depend on the sign of $\dot\rho$).

This leaves the discussion of the background action and the boundary term of $G$. As mentioned before, the bounce in de Sitter does not end at $\phi_+$ but at some 
different value $\phi_f \not= \phi_+$. Furthermore, due to the sign changes one should split 
the integration ranges according to the sign of $\dot\rho$. For the function $G$ one then finds the contribution 
(notice that $G$ vanishes when $\dot\rho=0$)
\bea
\int_f^0 \frac{dG}{d\phi} d\phi &=& \int_f^\times \frac{dG}{d\phi} d\phi +\int_\times^0 \frac{dG}{d\phi} d\phi 
= \frac{12\pi^2}{\kappa^2} \left[ \frac{0}{V_\times} - \frac{1}{V(\phi_f)} \right] - \frac{12\pi^2}{\kappa^2} \left( \frac{1}{V_0} - \frac{0}{V_\times} \right) \nonumber\\ 
&=&
- \frac{12\pi^2}{\kappa^2} \left[ \frac{1}{V_0} + \frac{1}{V(\phi_f)} \right] \, ,
\eea
while the background contribution is 
\be
S_{bg} = -\frac{24\pi^2}{\kappa^2} \frac{1}{V_+} \, .
\ee
Overall, these terms give
\be
\left. G \right|_f^0 - S_{bg} = \frac{12 \pi^2}{\kappa^2} \left[ \frac{1}{V(\phi_f)} - \frac{1}{V_0} \right]
+ \frac{24 \pi^2}{\kappa^2} \left[ \frac{1}{V_+} - \frac{1}{V(\phi_f)} \right] \, .
\ee
The first term will again combine with (\ref{eq:lagfinal1}) to obtain the desired Lagrangian (\ref{eq:lagfinal2}).
The second piece can be naturally absorbed into the new action (\ref{eq:lagfinal2}) when the integration range is extended to include the interval from $\phi_f$ to $\phi_+$ with $V_t=V$, see~\cite{Espinosa:2018voj,Espinosa:2021tgx}.

\section{Concluding remarks}
\label{sec:concl}

In summary, the derivation of the new action involving the potential $V_t$ is quite transparent when using
canonical transformations. While the original derivation~\cite{Espinosa:2018voj} relied on reverse engineering 
the action from the equations of motion in the field space, the procedure here is much simpler and straightforward. 

The results of the $V_t$ formalism have some very interesting features. In particular, there is no reference to space-time. Moreover, in the case with gravity, while the original formulation involves two quantities (the radius $\rho$ and the scalar field value $\phi$), in the $V_t$ formulation only one quantity ($V_t$) encodes all the relevant information. This is due to the fact that the method obeys the Friedman law by construction. 
Moreover, no action from a background contribution has to be calculated and the result is 
automatically finite even if $\rho\to\infty$ for $\xi\to\infty$. The formalism also clarifies the relation between
the Coleman-de Luccia~\cite{Coleman:1980aw} and the Hawking-Moss bounces~\cite{Hawking:1981fz}. Most of these features have been discovered already in~\cite{Espinosa:2018voj}, but the canonical transformations provide further understanding and justification for them.

\vskip 1cm

One might wonder how the canonical transformation provides an action that 
is minimized by the bounce solution instead of merely extremized. What happened to the negative 
eigenmode that signals a tunneling process
in an unstable situation?
The reason seems to be linked to the fact the the canonical momentum and 
the canonical coordinate change roles in the Euclidean action. 

For simplicity consider the following Lagrangian of one single degree of freedom
in Euclidean time. By Euclidean we mean that the potential term changed sign and the 
Hamiltonian is unbounded from below.
\be
L = \frac12 \dot q^2 + \frac12 \lambda q^2 \, .
\ee
The corresponding Hamiltonian reads
\be
H = \frac12 p^2 - \frac12 \lambda q^2 \, ,
\ee
and, as mentioned, it is unbounded from below. Now consider the canonical transformation induced by 
the generating function $G(q,Q) = \sqrt{\lambda} \, q\, Q$. Then using the relations
\be
p = \frac{\partial G}{\partial q} = \sqrt{\lambda} \, Q \, , \quad
P = -\frac{\partial G}{\partial Q} = -\sqrt{\lambda} \, q \, , \quad
K = H \, ,
\ee
the new Hamiltonian reads
\be
K = -\frac12 P^2 + \frac12 \lambda Q^2 \, ,
\ee
and the new Lagrangian
\be
L' = -\frac12 \dot Q^2 - \frac12 \lambda Q^2 \, .
\ee
Obviously the sign of the Lagrangian changed. Two important ingredients that make this happen are, first,
the fact that the momenta and coordinates changed roles, and, second, that the 
sign change relies on the fact that the Lagrangian is Euclidean and the potential term is negative 
for the Hamiltonian. Both of these features are shared by the tunneling analysis when 
transforming to the new potential method. Notice that also in this example one has to add the boundary term $\left. G \right|$ in order to 
reproduce the original value for the action on-shell.

\vskip 1cm

Another interesting question is if the new potential method could help to calculate the functional determinant for the prefactor of the decay rate.
Depending on the model, every time the quantum one-loop contributions are 
relevant, also the functional determinant can have a sizeable impact on the tunneling probability. 

At first, this new $V_t$ route seems hopeless, since as mentioned before, the functional determinant in the new formulation 
is positive (the action is a minimum) while in the conventional approach it is not (there is a negative eigenmode related to decay).
However, the canonical transformation gives an explicit relation between the two actions that could
help to relate the two functional determinants. 

Still, there are several arguments that this connection might not be so easy to establish. 
For example, the new formulation 
is agnostic about spacetime. On the other hand, the functional determinant in the conventional 
approach incorporates this information [before the $O(4)$ symmetry is imposed] and leads to degeneracies in the functional determinant described by spherical harmonics. 

This point then intertwines with another aspect: the functional determinant is divergent due to 
(quantum) one-loop contributions and has to be regularized and renormalized. Moreover, a background 
contribution has to be subtracted to make the functional determinant finite. Here, the above mentioned 
degeneracies are required to make contact to renormalization in e.g.~scattering amplitudes.

Ultimately, the canonical transformation could help to establish an alternative way to understand these connections better and to calculate the relevant functional determinant from a formulation using $V_t$ as sole dynamical degree of freedom.

\section*{Acknowledgements}

The work of JRE and RJ is supported by the grants IFT Centro de Excelencia Severo Ochoa SEV-2016-0597, CEX2020-001007-S and by PID2019-110058GB-C22 funded by MCIN/AEI/10.13039/501100011033 and by ERDF.
TK is supported by the Deutsche Forschungsgemeinschaft (DFG, German Research Foundation) under Germany’s Excellence Strategy -- EXC 2121 “Quantum Universe” -- 390833306.

\appendix

\section{Alternative treatment of case with gravity}

In this appendix we reconsider the case with gravity for Minkowski or AdS false vacua. We show how the bounce and the background actions can be combined from the beginning leading to intermediate expressions that are finite and have $\kappa=0$ limits in agreement with the zero gravity case.

For the calculation of tunneling actions, as explained in the main text, the object of interest is 
the difference between the Euclidean action for the tunneling bounce (the Coleman-De Luccia instanton) and the Euclidean action for the false vacuum background
\be
\Delta S_E = S_E[\phi,\rho]-S_{bg}\ ,
\ee
where $S_{bg} \equiv S_E[\phi_+,\rho_+]$ with $\phi_+$ being the false vacuum field value and $\rho_+$ the corresponding metric function. Here $S_E[\phi,\rho]$ is as given in (\ref{SEgrav}), with $\xi_{\rm max}=\infty$.
Note that the equations of motion derived from $\Delta S_E$ are exactly the same as those from $S_E$ written above, as $S_{bg}$ is just a constant.

As discussed in the text, for the decay of Minkowski or AdS vacua, both actions are divergent at large $\xi$ and only its difference is finite. A convenient way to
deal with $\Delta S_E$ and the cancellation of divergences in these cases is to rewrite $S_E[\phi_+,\rho_+]$ as an integral in $\xi$ by mapping $\rho_+(\xi_+)=\rho(\xi)$ (which can always be done as both are monotonic functions defined over the same interval and reaching from 0 to $\infty$). This leads to the relation 
\be
\frac{d\xi_+}{d\xi}=\frac{\dot\rho}{\sqrt{1 - V_+ \kappa \rho^2/3}}\ .
\ee
Using it, we get
\bea
S_{bg} & =& 2\pi^2 \int_0^\infty
\left[ \rho^3 V_+ - \frac{3\rho}{\kappa} (2 - V_+ \kappa \rho^2/3) \right]\frac{\dot\rho}{\sqrt{1 - V_+ \kappa \rho^2/3}} \, d\xi \,
\nonumber\\
&=& \int_0^\infty d\xi \frac{d}{d\xi}\left[
\frac{12\pi^2}{V_+ \kappa^2}(1 - V_+ \kappa \rho^2/3)^{3/2}
\right] \nonumber\\
&=& -\int_{\phi_+}^{\phi_0} d\phi \frac{d}{d\phi}\left[
\frac{12\pi^2}{V_+ \kappa^2}(1 - V_+ \kappa \rho^2/3)^{3/2}
\right]\ .
\eea
The last expressions show explicitly that 
the integrand of this background action is a total derivative, compare with (\ref{Sbg}).

In order to use the canonical transformation, we write the tunneling action $\Delta S_E$ as a functional of 
$\xi(\phi)$ and $\rho(\phi)$, as done in the main text. This yields
\bea
\Delta S_E &=& -2\pi^2 \int^{\phi_0}_{\phi_+} \left\{
\left[ \rho^3 \left( \frac1{2 \xi'^2} + V\right)
- \frac{3\rho}{\kappa}\left(\frac{{\rho'}^2}{\xi'^2} + 1\right) \right] \, \xi' \right.\nonumber\\
&&\left.
-\frac{d}{d\phi}\left[
\frac{6(1 - V_+ \kappa \rho^2/3)^{3/2}}{V_+ \kappa^2}
\right]
\right\}d\phi\ ,
\eea
where the prime denotes, as usual, derivatives with respect to $\phi$.

To get the Hamiltonian of this system we first find the canonical momenta 
\be
p_\rho = \frac{\partial L}{\partial \rho'} = -\frac{12 \pi^2 \rho}{\kappa}\left(\sqrt{1 - V_+ \kappa \rho^2/3}-\frac{\rho'}{\xi'}\right)
\ee
and
\be
p_\xi = \frac{\partial L}{\partial \xi'} = - \frac{\rho \pi^2}{\kappa \xi'{}^2} 
\left[ 6(\rho'^2 - \xi'^2) + \kappa \rho^2 (2 V \xi'^2 - 1) \right] \, .
\ee
The Legendre transformation gives the Hamiltonian 
\be
H = \frac{2 \pi^2 \rho^3}{\xi'} \, ,
\ee
which is the same as (\ref{Hgrav}) when expressed in these variables. As in the main text, $p_\xi\equiv 0$ during the evolution
and, in terms of the canonical variables, one has
\be
H = - \rho \sqrt{p_\rho^2\kappa/6 + 4 \pi^2 p_\rho \rho\sqrt{1 - V_+ \kappa \rho^2/3} + 8 \pi^4 (V-V_+) \rho^4} \, ,
\ee
which is somewhat more complicated than the Hamiltonian found in the derivation of the main text.

Introducing as usual the tunneling potential as
\be
V_t = V- \frac{1}{2 \xi'^2} =V_+ - \frac{p_\rho^2 \kappa}{48 \pi^4 \rho^4} - \frac{p_\rho\sqrt{1 - V_+ \kappa \rho^2/3}}{2\pi^2 \rho^3} \, ,
\ee
we express $p_\rho$ in terms of $V_t$ and $\rho$ as
\be
p_\rho = 12 \pi^2 \frac{\rho}{\kappa} \left[\sqrt{1 - V_t \kappa \rho^2/3} -\sqrt{1 - V_+ \kappa \rho^2/3}\right]\, ,
\ee
which is readily integrated to obtain the generating function
\be
G = -\frac{12\pi^2}{\kappa^2}
\left[\frac{
 (1- V_t \kappa \rho^2/3)^{3/2}-1}{V_t}- \frac{
 (1- V_+ \kappa \rho^2/3)^{3/2}-1}{V_+}
 \right]\, .
\ee
This $G$ has several good properties. First, its small $\kappa$ expansion gives
\be
G=-\frac{\pi^2}{2}(V_t-V_+)\rho^4+ {\cal O}(\kappa)\ ,
\ee
which reduces to the case without gravity [see (\ref{G0}) with $V_+=0$]. Second, $G$ vanishes both at $\rho=0$ and $\rho=\infty$ and we do not have to bother about the contribution $\left. G \right|$.

From $G$ above we get the canonical momentum 
\be
P = -\frac{\partial G}{\partial V_t} = -\frac{12\pi^2}{\kappa^2 V_t^2}
\left[\left(1 + V_t \kappa \rho^2/6\right) \sqrt{1 - V_t \kappa \rho^2/3}-1\right] \, ,
\ee
and the Hamiltonian 
\be
K = -2 \pi^2 \rho^3 \sqrt{2(V - V_t)}\ ,
\ee
which is the same as that found in (\ref{K}).

To get the action for the tunneling potential we need $V_t'$, which can be found easily via the chain rule as
\be
V_t' = \frac{\partial K}{\partial P} = \frac{3 K}{\rho} \left(\frac{\partial P}{\partial \rho}\right)^{-1}
= -\frac{3}{\rho} \sqrt{2(V-V_t)} \sqrt{1- V_t \kappa \rho^2/3} \, .
\ee
Using also $D^2 = V_t'^2 + 6 \kappa (V - V_t)V_t = 18(V - V_t)/\rho^2$, the final Lagrangian reads
\be
L = \frac{6\pi^2}{\kappa^2} \frac{(D + V_t')^2}{D V_t^2} \, .
\ee
This expression is, again, the one obtained in the original derivation~\cite{Espinosa:2018voj}.

\newpage

\bibliographystyle{JHEP}
\bibliography{refs}

\end{document}